\documentclass{article}
\usepackage{authblk}
\usepackage[utf8]{inputenc}
\usepackage{parskip}
\usepackage{textcomp}
\usepackage{comment}
\usepackage[sorting=none]{biblatex}
\usepackage{graphicx}
\usepackage{epsfig}
\usepackage{amsmath}
\usepackage{subcaption}
\graphicspath{ {./image/} }
\usepackage{amssymb}
\usepackage{float}
\providecommand{\keywords}[1]
{
  \small	
  \textbf{\textit{Keywords---}} #1
}

\usepackage{hyperref}
\hypersetup{
    colorlinks=true,
    linkcolor=blue,
    filecolor=magenta,      
    urlcolor=cyan,
    pdftitle={Overleaf Example},
    pdfpagemode=FullScreen,
    }
\usepackage{geometry}
\begin{document}
\date{}
\title{Thermodynamics of BTZ Black Holes in Nonlinear Electrodynamics}

\author[1]{Prosenjit Paul\thanks{prosenjitpaul629@gmail.com}}
\author[2,3]{S. I. Kruglov\thanks{serguei.krouglov@utoronto.ca}}
\affil[1]{Indian Institute Of Engineering Science and Technology (IIEST), Shibpur-711103, WB, India}
\affil[2]{Department of Physics, University of Toronto, \protect\\ 60 St. Georges St.,
Toronto, ON M5S 1A7, Canada} 
\affil[3]{Canadian Quantum Research Center, 204-3002 32 Ave Vernon, BC V1T 2L7, Canada.}

\maketitle
\begin{center}
\rule{17cm}{0.4mm}
\end{center}

\begin{abstract}
We investigate electrically charged black holes in $(2+1)$ dimensional gravity coupled to nonlinear electrodynamics (NED). The metric function $f(r)$ is depicted, showing that there can be one or two horizons. We study black hole thermodynamics in extended phase space. The cosmological constant played the role of thermodynamics pressure. Electrostatic potential, vacuum polarization, internal energy, Helmholtz and Gibbs free energies as functions of black hole horizon and Hawking temperature were analysed. The global stability of the black hole is discussed. The specific heat and local thermodynamic stability of black holes are studied. Finally, the Van der Walls behaviour of the black holes is investigated.
\end{abstract}
\newpage
\begin{center}
\rule{17cm}{0.4mm}
\end{center}
\tableofcontents
\begin{center}
\rule{17cm}{0.4mm}
\end{center}

\section{Introduction}
Black holes are fascinating objects in our observable universe. The discovery that black holes behave as thermodynamic systems has changed our understanding of general relativity and its relationship to quantum field theory. The entropy of a black hole is proportional to the area of the event horizon \cite{bekenstein1973black}. The work of Bardeen, Carter and Hawking showed four laws of black hole thermodynamics \cite{bardeen1973four}. Taking account into relativistic quantum effects Hawking showed \cite{hawking1974black,hawking1975particle} that black holes emit thermal radiation at a temperature inversely proportional to the mass of the black hole. Hawking and Page found \cite{hawking1983thermodynamics} a solution of Einstein's field equation with a negative cosmological constant. 

The discovery of the famous BTZ black hole \cite{banados1992black,banados1993geometry} attracted much attention to study $(2+1)$ dimensional black holes. Nowadays, black holes in $(2+1)$ dimensions were studied in great detail, e.g. thermodynamics of $(2+1)$ dimensional rotating AdS black hole \cite{zaslavskii1994thermodynamics}, charged rotating solution was obtained in Ref. \cite{martinez2000charged}, thermodynamics of $(2+1)$ dimensional black hole \cite{frassino2015lower}, static charged black holes in dilaton gravity \cite{chan1994static}, black holes in  Einstein--Born--Infeld--dilaton theory \cite{yamazaki2001black}, magnetic black hole in Einstein-Maxwell theory with negative cosmological constant \cite{hirschmann1996magnetic}, electrically charged black holes in Einstein-power-Maxwell theory \cite{gurtug20122+}. Some aspects of three-dimensional black holes were considered in Refs. \cite{Koch:2016uso,Rincon:2017goj}. Here, we study $(2+1)$ dimensional electrically charged black hole in a NED model proposed in \cite{kruglov2017nonlinear} (see also Ref. \cite{mangut2020quantum}), coupled to gravity with $AdS$ spacetime.

The NED models have been developed with the aim of resolving singularities and incorporating quantum gravity corrections into classical Maxwell fields. Interestingly, the weak-field limit of NED models recovers the well-known Maxwell fields, making NED a valuable addition to the study of electromagnetic fields. Another interesting feature of NED is that electric fields due to a point-like charge are finite at the origin. Born-Infeld electrodynamics \cite{Born:1934gh} is the first known example of NED. Black holes in $3D$ Einstein gravity coupled to Born-Infeld electrodynamics studied in Refs. \cite{Cataldo:1999wr,Myung:2008kd}. Black hole solution in $4D$ Einstein gravity coupled to Born-Infeld electrodynamics studied in Ref \cite{Gunasekaran:2012dq}. The interest in the model \cite{Kruglov:2022lnc} is due to its simplicity. The mass and metric functions are expressed via elementary functions. The NED model \cite{kruglov2017nonlinear,Kruglov:2022lnc} under consideration possesses similar characteristics that bear resemblance to those of Born-Infeld electrodynamics \cite{Born:1934gh}. The $4D$ Einstein gravity coupled to NED Lagrangian \cite{kruglov2017nonlinear} studied in Refs. \cite{kruglov2017nonlinear,Kruglov:2022lnc}. Black hole solutions in $3D$ Einstein gravity coupled to NED \cite{kruglov2017nonlinear} are not studied, which gives us an opportunity to fill this gap.

In this paper, we find a new electrically charged black hole solution in $(2+1)$ dimensional gravity coupled to NED. In section \ref{sec:2}, we study the solutions of Einstein field equation within the framework of NED and find the black hole solution. The metric function $f(r)$ for different values of $\beta$ (Fig. \ref{fig:1}a) and charges (Fig. \ref{fig:1}b) is depicted. In Fig. \ref{fig:1}b, the position of horizon increases as charges increase. In section \ref{sec:3}, we study the thermodynamics of black holes. We compute the Hawking temperature of black holes and derive first law of black hole thermodynamics in extended phase space. The thermodynamical parameters (e.g. internal energy, Helmholtz free energy and Gibbs free energy) associated with a black hole also are studied.  Finally, we compute the specific heat of the black hole and plot the specific heat for different values of $\beta$ (Fig. \ref{fig:6}a) and charges (Fig. \ref{fig:6}b). It is seen from the plot that no phase transition occurs for electrically charged black holes in $(2+1)$ dimensional gravity coupled to NED. In section \ref{sec:4}, we study the Van der Waals-like phase transition of the black holes and it is found that no positive solutions are possible for critical temperature and pressure.

\section{NED-\texorpdfstring{$AdS$}{TEXT} black hole solution}\label{sec:2}

In this section, we study the $(2+1)$ dimensional Einstein gravity coupled to NED in $AdS$ background. We find the gravitational and electromagnetic field equations. Finally, the metric function and the electric field of point-like charges are obtained. The action for $(2+1)D$ Einstein gravity coupled to  NED in $AdS$ background is given by

\begin{equation}\label{eq:2.1}
I = \int d^{3}x \sqrt{{-g}} \biggr[\frac{R-2\Lambda}{16 \pi} + \mathcal{L}(F)     \biggr], 
\end{equation}

where $\Lambda$ is cosmological constant and $\Lambda = {-1}/{l^2} $, $l$ is the $AdS$ radius, with the Newton constant $G=1$. The  NED Lagrangian $\mathcal{L}(F)$ is given by \cite{kruglov2017nonlinear}

\begin{equation}\label{eq:2.2}
 \mathcal{L}(F) = -\frac{F}{1 + \epsilon \sqrt{ \lvert \beta F \rvert }},   
\end{equation}

with 
\begin{equation}\label{eq:2.3}
    F=  \frac{1}{4} F_{\mu \nu} F^{\mu \nu},
\end{equation}

where $\beta$ has the dimensions of $(length)^{3}$,  $\epsilon=\pm 1$ and $F_{\mu \nu} = \partial_{\mu}A_{\nu}  - \partial_{\nu}A_{\mu}$. In the limit $\beta \to 0$ NED Lagrangian reduces to Maxwell's Lagrangian. The electric and magnetic fields are defined as $F_{tr} = E$ and $F_{\phi r} = B $. Varying action \eqref{eq:2.1} with respect to $g_{\mu \nu}$ and ${A_{\mu}}$, we get the gravitational and electromagnetic field equations

\begin{equation}\label{eq:2.4}
     G_{\mu \nu} + \Lambda g_{\mu \nu} = 8 \pi T_{\mu \nu},
\end{equation}
\begin{equation}\label{eq:2.5}
    \nabla_{\mu} \biggr[\sqrt{-g} \Bigl( \partial_{F} \mathcal{L}(F)\Bigl)    F^{\mu \nu}  \biggr] = 0,
\end{equation}

where $T_{\mu \nu}=  \mathcal{L}(F ) g_{\mu \nu} - F_{\mu \alpha} F_{\nu}^{\alpha} \partial_{F} \mathcal{L}(F)$ and $\partial_{F} \mathcal{L}(F) = \partial \mathcal{L}(F) /\partial {F} $. The magnetically and electrically charged black hole solutions in fourth dimensions with the Lagrangian \eqref{eq:2.2} were found in \cite{kruglov2017nonlinear,mangut2020quantum}. Here we consider the static and spherically symmetric metric

\begin{equation}\label{eq:2.6}
     ds^{2} = -  f(r) dt^{2} + \frac{dr^{2}}{f(r)} + r^2 d{\phi}^{2}.
\end{equation}

The magnetic field vanishes and only electric field plays the role of the gravitational source \cite{cataldo2000regular}. To find the electric field we will consider the electromagnetic field equation \eqref{eq:2.5}

\begin{equation}\label{eq:2.7}
    \partial_{r} \Biggr[r E\frac{(1+ \frac{\epsilon aE}{2})}{(1+ {\epsilon aE})^2}   \Biggr]=0,
\end{equation}

with the solution 
\begin{equation}\label{eq:2.8}
    E=- \frac{1}{\epsilon a} \Biggr[ 1 \pm \sqrt{\frac{r}{r-2\epsilon aq}} \Biggr],
\end{equation}

where $a=\sqrt{{\beta}/{2}}$, $q$ is the integration constant and it is related to the charge of black holes. We choose negative branch of the electric field solution because in the limit $\beta \to 0$ negative branch gives the correct solutions for electric field. Therefore, the physical solution of the electric field is given by 

\begin{equation}\label{eq:2.9}
E=\frac{1}{a} \Biggr[ 1 - \sqrt{\frac{r}{r+2 aq}} \Biggr],
\end{equation}

where we choose $\epsilon=-1$ and discard $\epsilon=1$, because  $\epsilon=1$ gives complex electric field when $r < 2aq$. In the limit $\beta \to 0$ electric field \eqref{eq:2.9} becomes

\begin{equation}\label{eq:2.10}
    E= \frac{q}{r}.
\end{equation}

Now considering the $\phi \phi$ component of Einstein equation \eqref{eq:2.4} we have

\begin{equation}\label{eq:2.11}
 R_{\phi\phi} = 
8 \pi  T_{\phi\phi} +2 (-8 \pi  T +\Lambda ) g_{\phi\phi},
\end{equation}

Substituting $R_{\phi \phi} = -r \frac{df(r)}{dr}$ into equation \eqref{eq:2.11} we get

\begin{equation}\label{eq:2.12}
    \frac{df(r)}{dr}= {-2 \Lambda r  +16 \pi r ( \mathcal{L(F)} + E^2 {\partial_{F}{\mathcal{F}}})}.
\end{equation}

Replacing the electric field from equation \eqref{eq:2.9} and integrating both sides of above equation one finds

\begin{equation}\label{eq:2.13} 
  f(r)= -M_{c} -\Lambda  r^{2} -\frac{4 q r}{a} +2 C r^{2}+\frac{2 (a q +r )}{a^{2}}  \sqrt{2 a q r +r^2}-2 q^{2} \ln  (a q +r +\sqrt{2 a q r +r^{2}}),
\end{equation}

where $M_c$ and $C$ are integration constants. We will adjust the constant $C$ according to our theory. We choose both constants in such a way that in the limit $\beta \to 0$ the metric function reduces to Maxwell-BTZ metric function \cite{martinez2000charged,frassino2015lower}. In the limit $\beta \to 0$, metric function becomes
\begin{equation}\label{eq:2.14}
    f(r)= -M + \frac{r^2}{l^2} -\frac{Q^2}{2} \ln (r/l),
\end{equation}

where $M_c=M+ 2 q^2 \ln ( e/2l)$, $C= -1/a^2$. $Q$ is the charge of black holes and it is related to the integration constant $q$ by the equality $Q=2q$. In the case of Maxwell electrodynamics we have $C=0$ and for Born--Infeld electrodynamics $C=b$ \cite{cataldo2000regular}. If we set the cosmological length equal to unity into equation \eqref{eq:2.14}, then the metric function of Maxwell electrodynamics reduces to $f= -M + {r^2} -({Q^2}/{2}) \ln (r)$, which was obtained in Ref. \cite{martinez2000charged} in the limit $J \to 0$ \& $Q \neq 0$.  Now, the equation \eqref{eq:2.13} takes the following form
\begin{equation}\label{eq:2.15}
    f(r)=-M +\frac{r^{2}}{l^2}-\frac{2 Q r}{a}- \frac{2  r^{2}}{a^2}+\frac{ (Qa +2r )\sqrt{Qar +r^2} }{a^{2}}  - \frac{Q^{2}}{2} \ln  \Biggr[\frac{e \Bigl(\frac{Qa}{2} +r +\sqrt{Qar  +r^{2}} \Bigl)}{2l} \Biggr].
\end{equation}

\begin{figure}[H]
\centering
\subfloat[$M=1$, $q=3$]{\includegraphics[width=.5\textwidth]{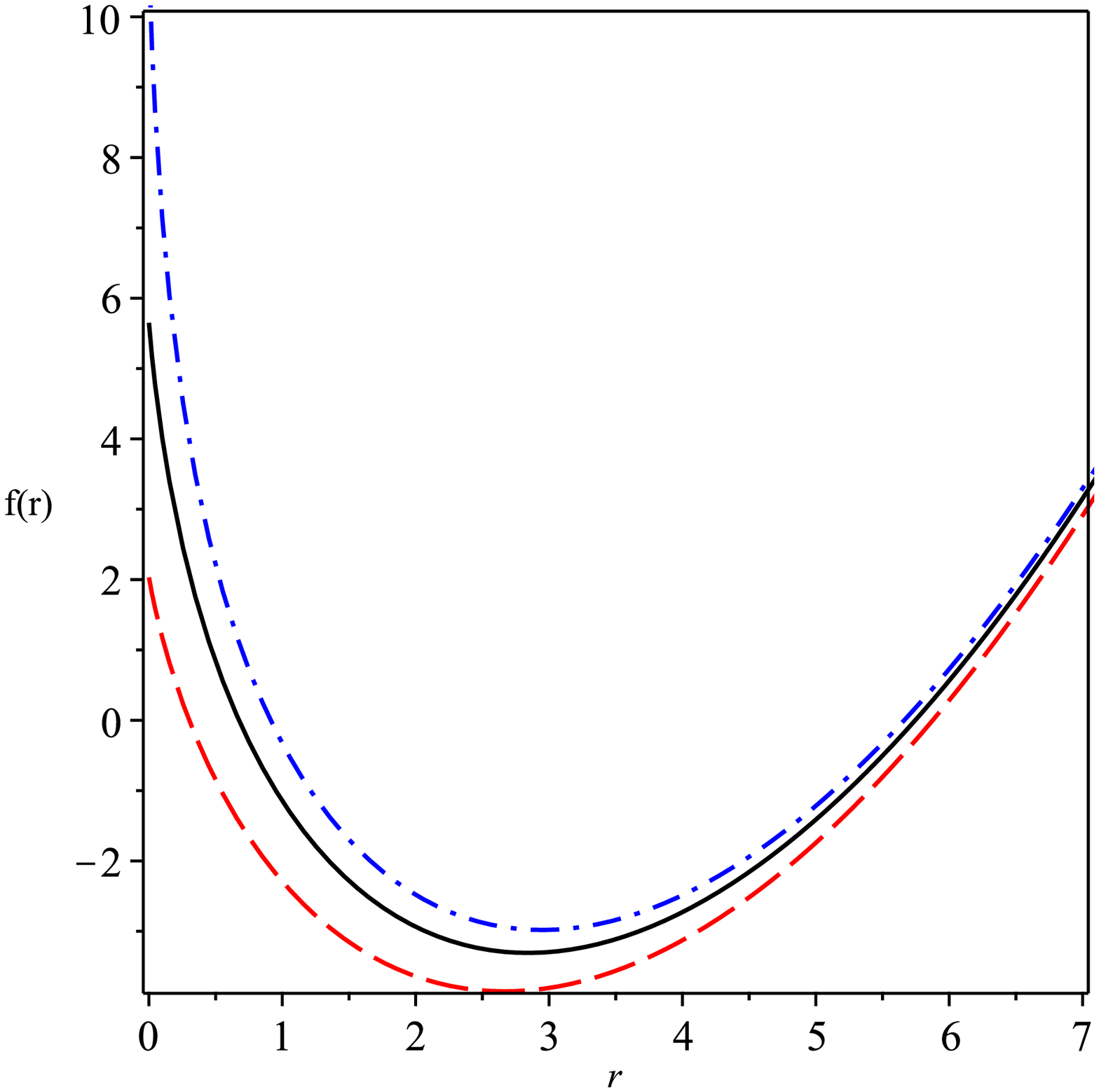}}\hfill
\subfloat[$M=1$, $\beta=1$]{\includegraphics[width=.5\textwidth]{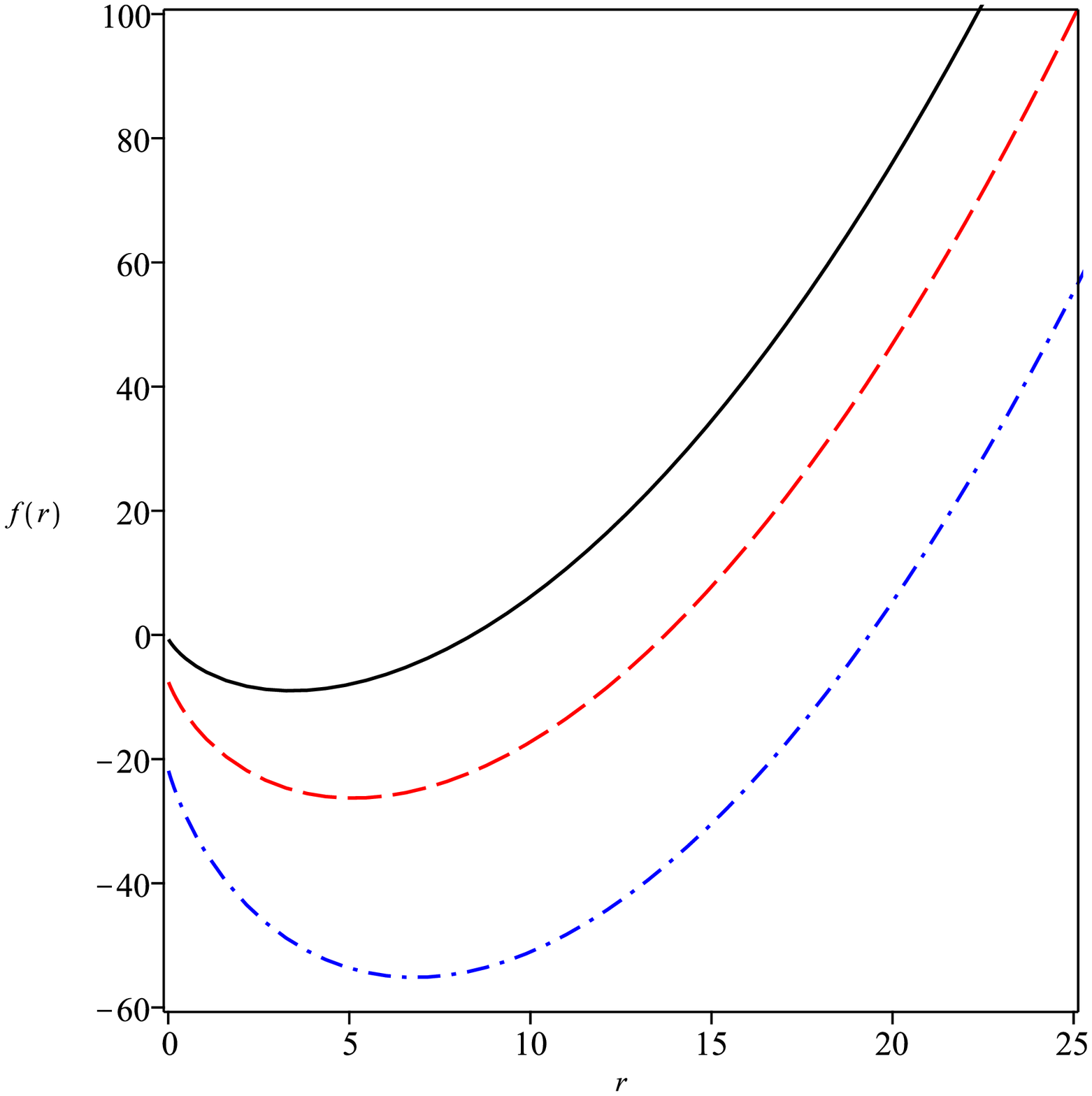}}\hfill
\caption{ Left panel: $l=2$ with $\beta=0.1$ denoted by black solid line, $\beta=0.5$ denoted by red dash line and $\beta=0.01$ denoted by blue dash-dot line.  Right panel: $l=2$ with $q=4$ denoted by black solid line, $q=6$ denoted by red dash line and $q=8$ denoted by blue dash-dot line.}\label{fig:1}
\end{figure}

In Fig. \ref{fig:1}(a) and Fig. \ref{fig:1}(b) function $f(r)$ is depicted for different values of $\beta$ and $Q$. In Fig. \ref{fig:1}(b), the effects of charges on the black hole horizon are shown, as charges of the black holes increase the position of the horizon also increases.

\section{Black hole thermodynamics}\label{sec:3}

In this section, we study black hole thermodynamics. We derive the Hawking temperature of black holes, first law of black hole thermodynamics and various thermodynamics parameters associated with black holes. To obtain the first law of black hole thermodynamics we need to introduce pressure, which is associated with the negative cosmological constant \cite{kubizvnak2012p,dolan2010cosmological}. The first law of black hole thermodynamics is $ dM=T_{H} dS + \Omega dJ + \Phi dq $ where $M$, $q$ and $J$ are the mass, charge and angular momentum of black holes. Including the pressure-volume term the first law of black hole thermodynamics takes the following form \cite{cvetivc2011black,kastor2009enthalpy,dolan2010cosmological} $ dM=T_{H} dS + V dP+ \Omega dJ + \Phi dq$. We define entropy and pressure of black hole

\begin{equation*}
     S=\frac{\pi r_{h}}{2}, 
\end{equation*}

\begin{equation}\label{eq:3.1}
    P=\frac{1}{8 \pi l^2}.
\end{equation}

At the position of event horizon $r_h$ ($f(r_h)=0$) one has

\begin{equation}\label{eq:3.2}
M= \frac{r_{h}^{2}}{l^2}-\frac{2 Q r_{h}}{a}- \frac{2  r_{h}^{2}}{a^2}+\frac{ (Qa +2r_{h} )\sqrt{Qar_{h} +r_{h}^2} }{a^{2}}  - \frac{Q^{2}}{2} \ln  \Biggr[\frac{e \Bigl(\frac{Qa}{2} +r_{h} +\sqrt{Qar_{h}  +r_h^{2}} \Bigl)}{2l} \Biggr].
\end{equation}

Hawking temperature of the black hole is defined as

\begin{equation}\label{eq:3.3}
  T_{H} = \frac{1}{4 \pi} \biggr[ \frac{df(r)}{dr} \biggr]_{r=r_{h}}.
\end{equation}

From the metric function \eqref{eq:2.15} and using equation \eqref{eq:3.3}, we obtain the Hawking temperature of black holes as

\begin{equation}\label{eq:3.4}
    T_{H} = \frac{1}{4\pi} \Biggr[ \frac{2 r_{h}}{l^{2}}-\frac{4 r_{h}}{a^{2}}-\frac{2 Q}{a}+\frac{4 \sqrt{r_{h}} \sqrt{Q a +r_{h}}}{a^{2}} \Biggr].
\end{equation}

Hawking temperature of rotating Maxwell-BTZ black hole was obtained in \cite{banados1992black}. If we expand the square root term in the above equation up to second order in $a$, then Hawking temperature of Maxwell-BTZ black hole \cite{frassino2015lower} is obtained. From  the equation \eqref{eq:3.2}, first law of black hole thermodynamics and generalised Smarr formula are obtained 

\begin{equation*}
    d\mathcal{M} = T_{H} dS +V dP + \Phi dQ + \mathcal{B} d{\beta},
\end{equation*}
\begin{equation}\label{eq:3.5}
    2PV=TS+ \mathcal{B} \beta,
\end{equation}

where $\Phi$ and $\mathcal{M}$ are the electrostatic potential and mass of black holes. The thermodynamic conjugate to NED parameter $\beta$ is $\partial{\mathcal{M}}/\partial{\beta}= \mathcal{B}$,  with

\begin{equation}\label{eq:3.6}
    \mathcal{M}= \frac{M}{8},
\end{equation}
\begin{equation}\label{eq:3.7}
V=\Biggl( \frac{\partial{\mathcal{M}}}{\partial{P}}\Biggl)_{S,Q,\beta}=  \pi r_{h}^2 - \frac{Q^2}{32P},
\end{equation}

\begin{equation*}
\Phi = -\frac{r_{h}}{4 a}+\frac{\sqrt{r_{h}} (3 Q a +4 r_{h} )}{16 \sqrt{Q a +r_{h}} a}-\frac{Q^{2} a ( \sqrt{Q ar_{h} +r_{h}^2}+r_{h} )}{\sqrt{Q ar_{h} +r_{h}^2} (16 Q a +32  \sqrt{Q ar_{h} +r_{h}^2}+32 r_{h} )}
\end{equation*}
\begin{equation}\label{eq:3.8}
-\frac{Q}{8} \ln \Biggr[e  \biggl(\frac{Q a}{2}+r_{h} +\sqrt{Q a r_{h} +r_{h}^{2}} \biggl)  \sqrt{2 \pi  P}\Biggr].
\end{equation}

\begin{figure}[H]
    \centering
    \includegraphics[scale=0.5]{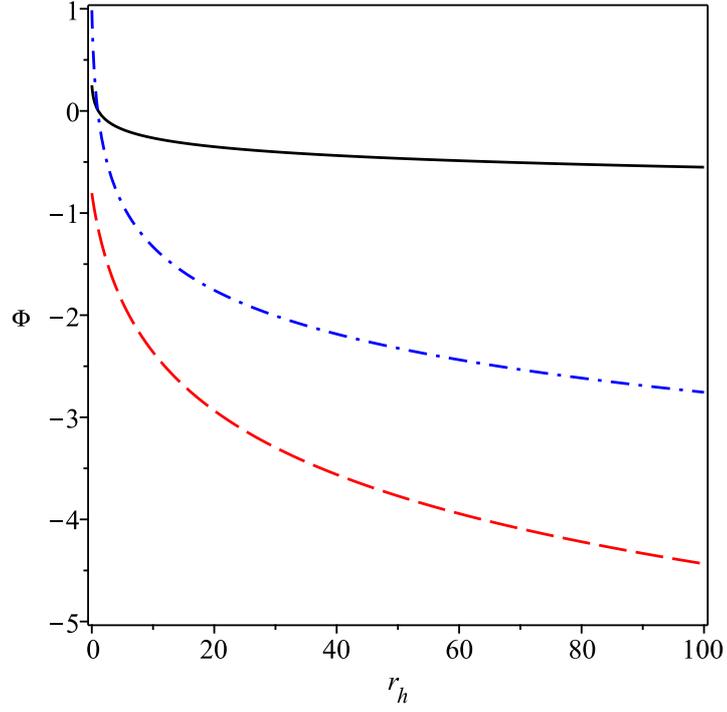}
    \caption{Electrostatic potential Vs black hole horizon with $l=2$, $\beta=0.1$ and $Q=1$ denoted by black solid line, $\beta=0.5$ and $Q=8$ denoted by red dash line, $\beta=0.01$ and $Q=5$ denoted by blue dash-dot line.}
    \label{fig:2}
\end{figure}

In the limit $\beta \to 0$ potential takes the following form \cite{frassino2015lower}
\begin{equation}
    \Phi=-\frac{Q}{8} \ln(r_h/l) +K,
\end{equation}
where $K=-Q/8 \ln(2e^2)$ is a constant. The constant $K$ does not affect our physical electric field. In Fig. \ref{fig:2} we plot the electrostatic potential for different values of $\beta$ and charges of the black holes. The figure shows that electrostatic potential is a decreasing function of the horizon radius and for a particular value of the horizon radius, the electrostatic potential is zero.

\begin{figure}[H]
    \centering
    \includegraphics[scale=0.5]{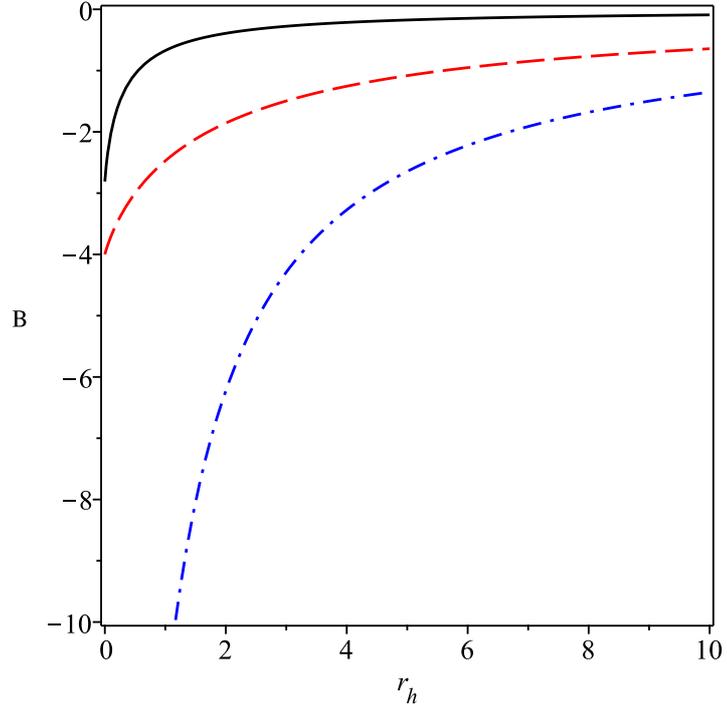}
    \caption{Vacuum polarization Vs black hole horizon with  $l=2$, $\beta=0.1$ and $Q=3$ denoted by black solid line, $\beta=0.5$ and $Q=8$ denoted by red dash line,  $\beta=0.01$ and $Q=5$ denoted by blue dash-dot line.}
    \label{fig:3}
\end{figure}

The concept of vacuum polarization in NED black holes was first studied in Ref. \cite{Gunasekaran:2012dq} for Born-Infeld electrodynamics. In quantum field theory, a vacuum contains short-lived virtual particle-antiparticle pairs. In the presence of a background electric field, these particle-antiparticle pairs reposition themselves and interact with the field. Vacuum polarization refers to the reorientation of particle-antiparticle pairs. The Vacuum polarization is necessary for the consistency of Smarr relation \eqref{eq:3.5}. In Fig. \ref{fig:3}, vacuum polarization for different values of $\beta$ and charges of black holes are depicted. Vacuum polarization starts from a negative value, then approaches zero, i.e. vacuum polarization is an increasing function of the horizon radius.

The vacuum polarization is defined as
\begin{equation}
    \mathcal{B}= \biggl( \frac{\partial \mathcal{M}}{\partial \beta} \biggl).
\end{equation}

\begin{equation}\label{eq:3.9}
\mathcal{B} = -\frac{Q^{2} \biggl( (\frac{Q a}{2}-r_{h} ) \sqrt{Q a +r_{h}}+Q a \sqrt{r_{h}}+r_{h}^{\frac{3}{2}} \biggl)}{32 a^{2} \sqrt{Q a +r_{h}}  \biggl( Q a +2 \sqrt{r_{h}} \sqrt{Q a +r_{h}}+2 r_{h} \biggl)}.
\end{equation}

The internal energy of black holes is defined as $\mathcal{M} - P V$. Using equations \eqref{eq:3.1}, \eqref{eq:3.2} and \eqref{eq:3.7} one can obtain internal energy of black holes as

\begin{equation}\label{eq:3.10}
U = -\frac{r_{h}^{2}}{4 a^{2}}+\frac{Q^{2}}{32}-\frac{Q r_{h}}{4 a}+\frac{ \sqrt{Q a r_{h} +r_{h}^2} \Bigl( \frac{Q a}{2}+r_{h} \Bigl)}{4 a^{2}}-\frac{Q^{2} }{16} \ln \Biggr[ \frac{e \Bigl( \frac{Q a}{2}+r_{h} +\sqrt{Q a r_{h} +r_{h}^{2}}\Bigl) }{2l}\Biggr].
\end{equation}

\begin{figure}[H]
\centering
\subfloat[l=2]{\includegraphics[width=.5\textwidth]{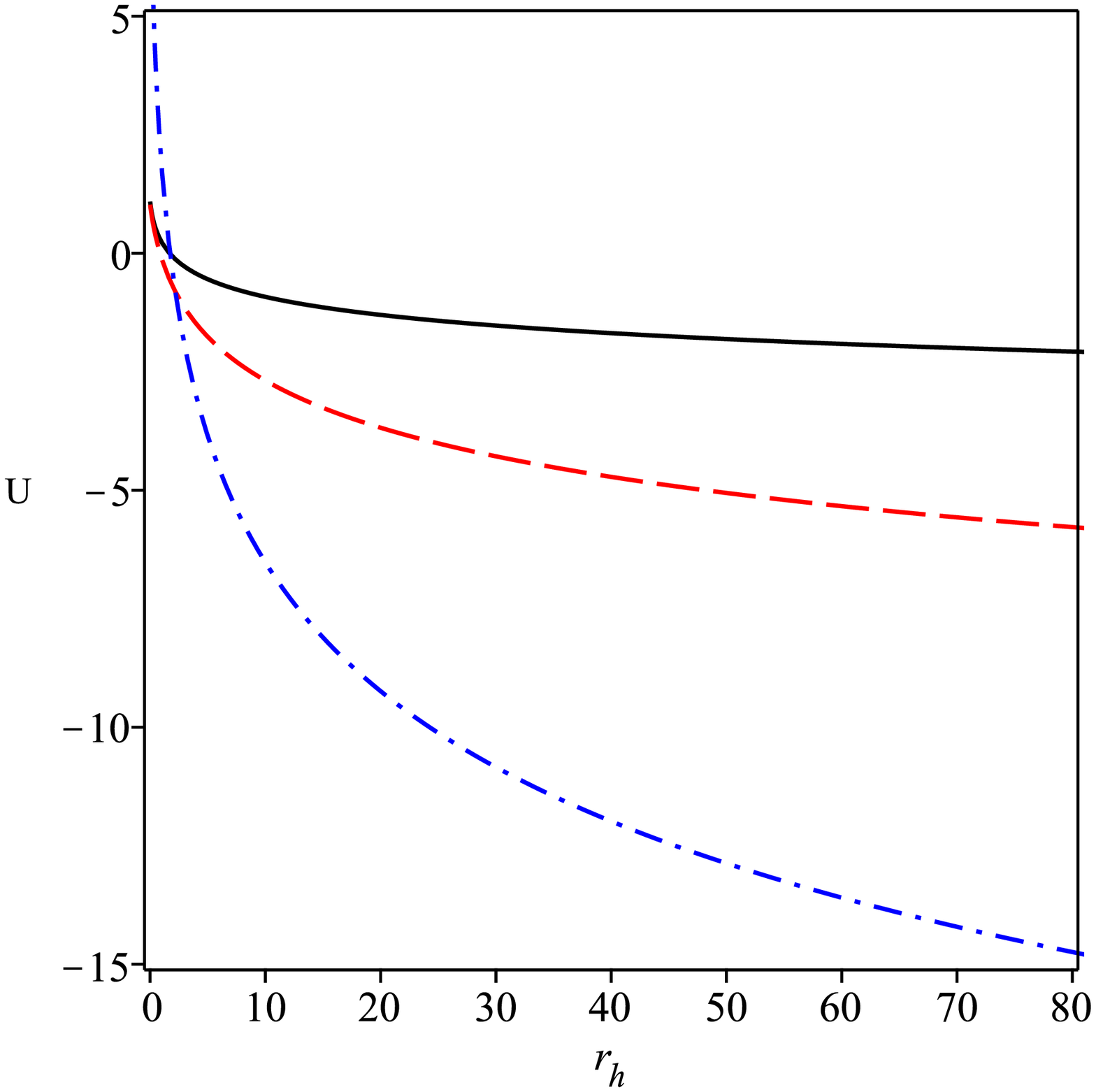}}\hfill
\subfloat[l=2]{\includegraphics[width=.5\textwidth]{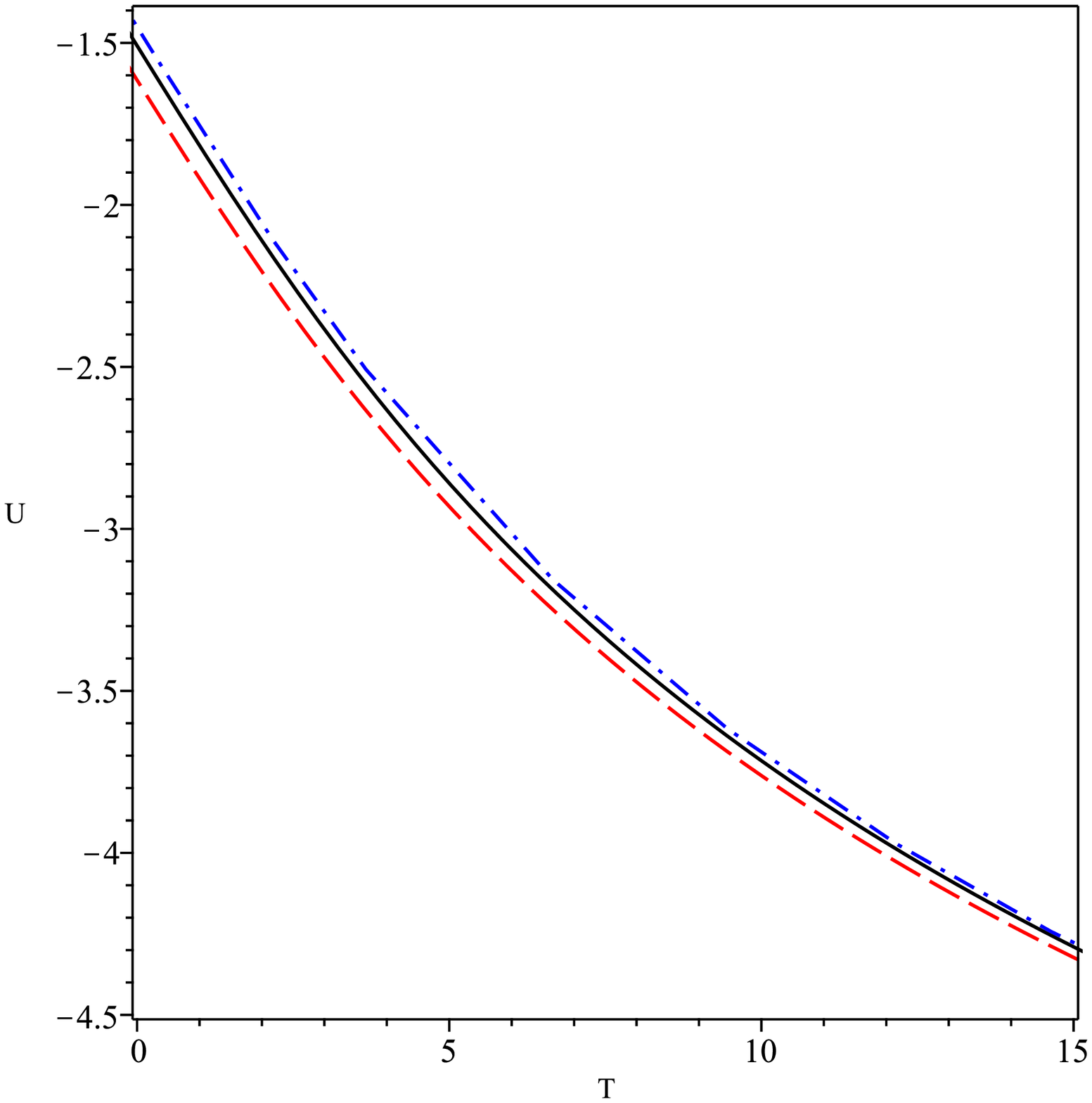}}\hfill
\caption{Left panel: Internal energy Vs black hole horizon. $\beta=0.1$ and $Q=3$ denoted by black solid line, $\beta=0.5$ and $Q=5$ denoted by red dash line, $\beta=0.01$ and $Q=8$ denoted by blue dash-dot line. Right panel: Internal energy Vs temperature with $Q=5$, $\beta=0.1$ denoted by black solid line, $\beta=0.5$ denoted by red dash line, $\beta=0.01$ denoted by blue dash-dot line.}\label{fig:4}
\end{figure}

The Helmholtz free energy is defined as 

\begin{equation}\label{eq:3.11}
    F= U- T_{H} S,
\end{equation}

and using equations \eqref{eq:3.1}, \eqref{eq:3.4} and \eqref{eq:3.10}, we obtain

\begin{equation*}
F=-\frac{r_{h}^{2}}{4 l^{2}}+\frac{r_{h}^{2}}{4 a^{2}} +\frac{Q^{2}}{32}-\frac{r_{h} \sqrt{Q ar_{h} +r_{h}^2}}{2 a^{2}}+\frac{ \sqrt{Q a r_{h}+r_{h}^2} \Bigl( \frac{Q a}{2}+r_{h}  \Bigl)}{4 a^{2}}
\end{equation*}  
\begin{equation}\label{eq:3.12}
-\frac{Q^{2}}{16} \ln \Biggr[\frac{e \Bigl( \frac{Q a}{2}+r_{h} +\sqrt{Q a r_{h} +r_{h}^{2}}\Bigl) }{2l} \Biggr].
\end{equation}    

\begin{figure}[H]
\centering
\subfloat[$l=2$]{\includegraphics[width=.5\textwidth]{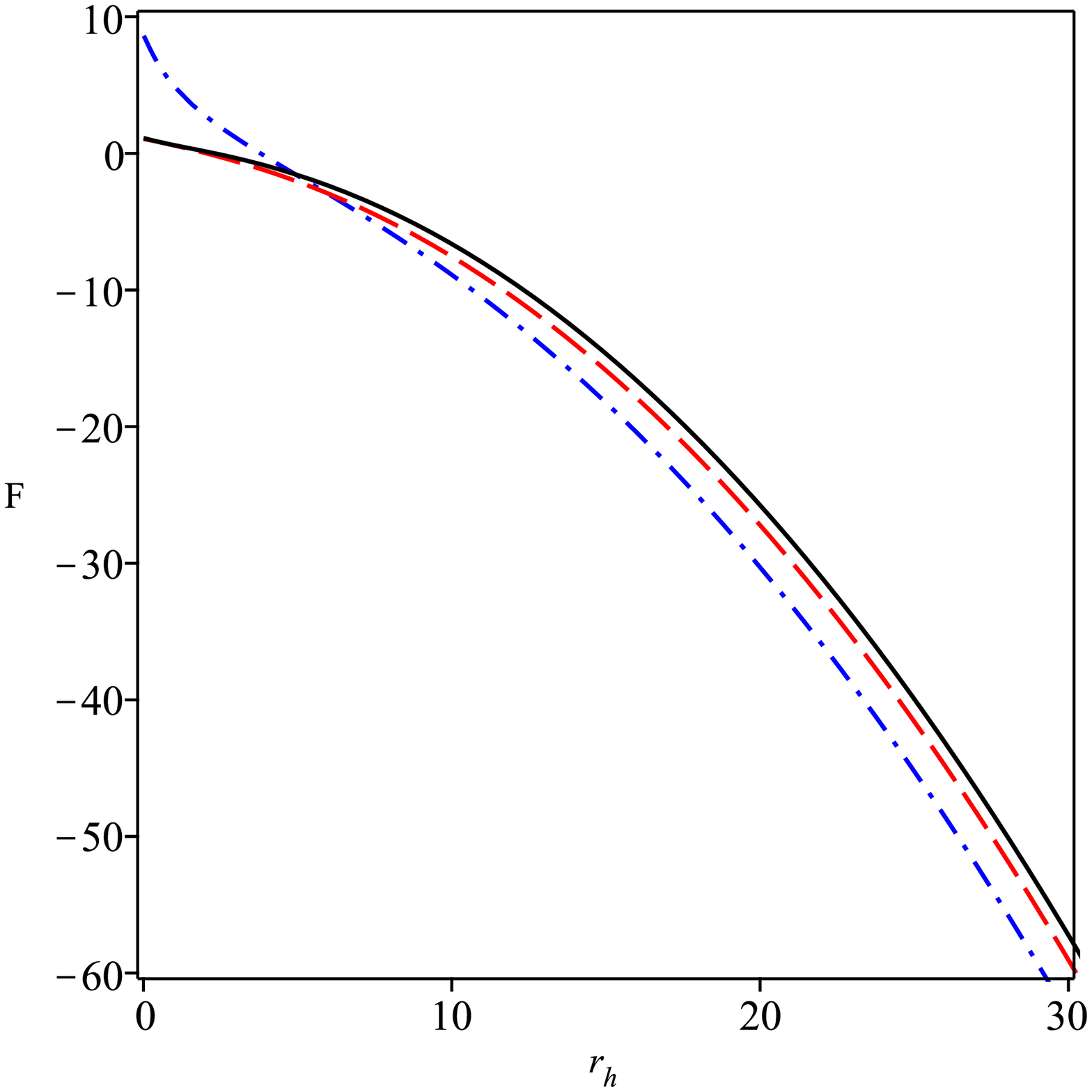}}\hfill
\subfloat[$l=2$]{\includegraphics[width=.5\textwidth]{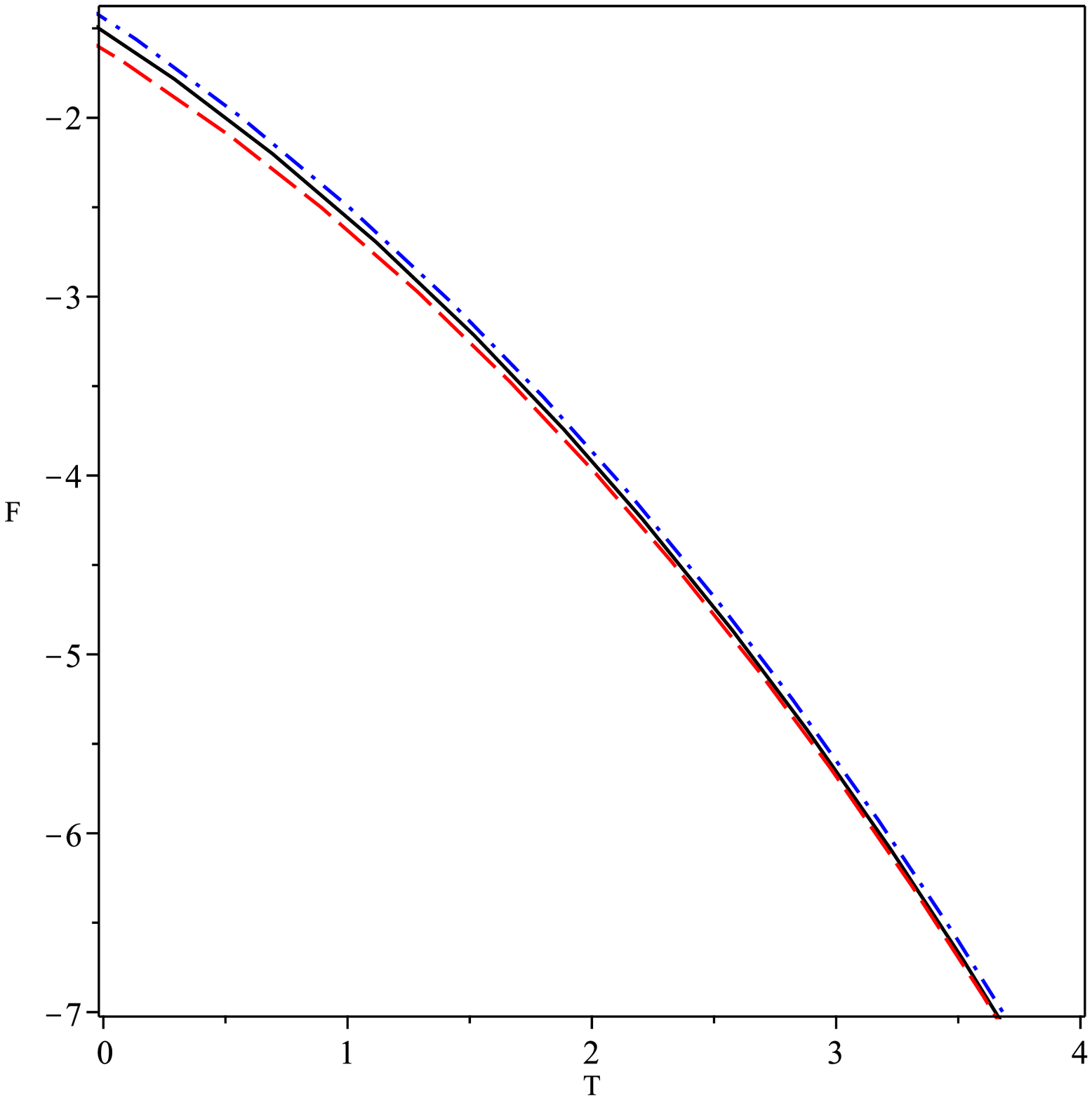}}\hfill
\caption{ Left panel: Helmholtz free energy Vs horizon with $\beta=0.1$ and $Q=3$ denoted by black solid line, $\beta=0.5$ and $Q=5$ denoted by red dash line, $\beta=0.01$ and $Q=8$ denoted by blue dash-dot line. Right panel: Helmholtz free energy Vs temperature  $\beta=0.1$ denoted by black solid line, $\beta=0.5$ denoted by red dash line \& $\beta=0.01$ denoted by blue dash-dot line with $Q=5$.}\label{fig:5}
\end{figure}

In Fig. \ref{fig:4}(a) and \ref{fig:4}(b), we plot the internal energy as a function of horizon radius and Hawking temperature. In Fig. \ref{fig:4}(a), internal energy starts from positive values and then decreases, for a critical value of the horizon radius internal energy is zero. In Fig. \ref{fig:4}(b), internal energy starts from negative values and then decreases, at $T=0$ internal energy is negative.  In Fig. \ref{fig:5}(a) and \ref{fig:5}(b), the Helmholtz free energy is depicted as a function of horizon radius and Hawking temperature. An increase in entropy corresponds to a decrease in Helmholtz's free energy \ref{fig:5}(a). At $T=0$  Helmholtz's free energy takes a negative value. 

\begin{figure}[H]
\centering
\subfloat[$Q=5$]{\includegraphics[width=.5\textwidth]{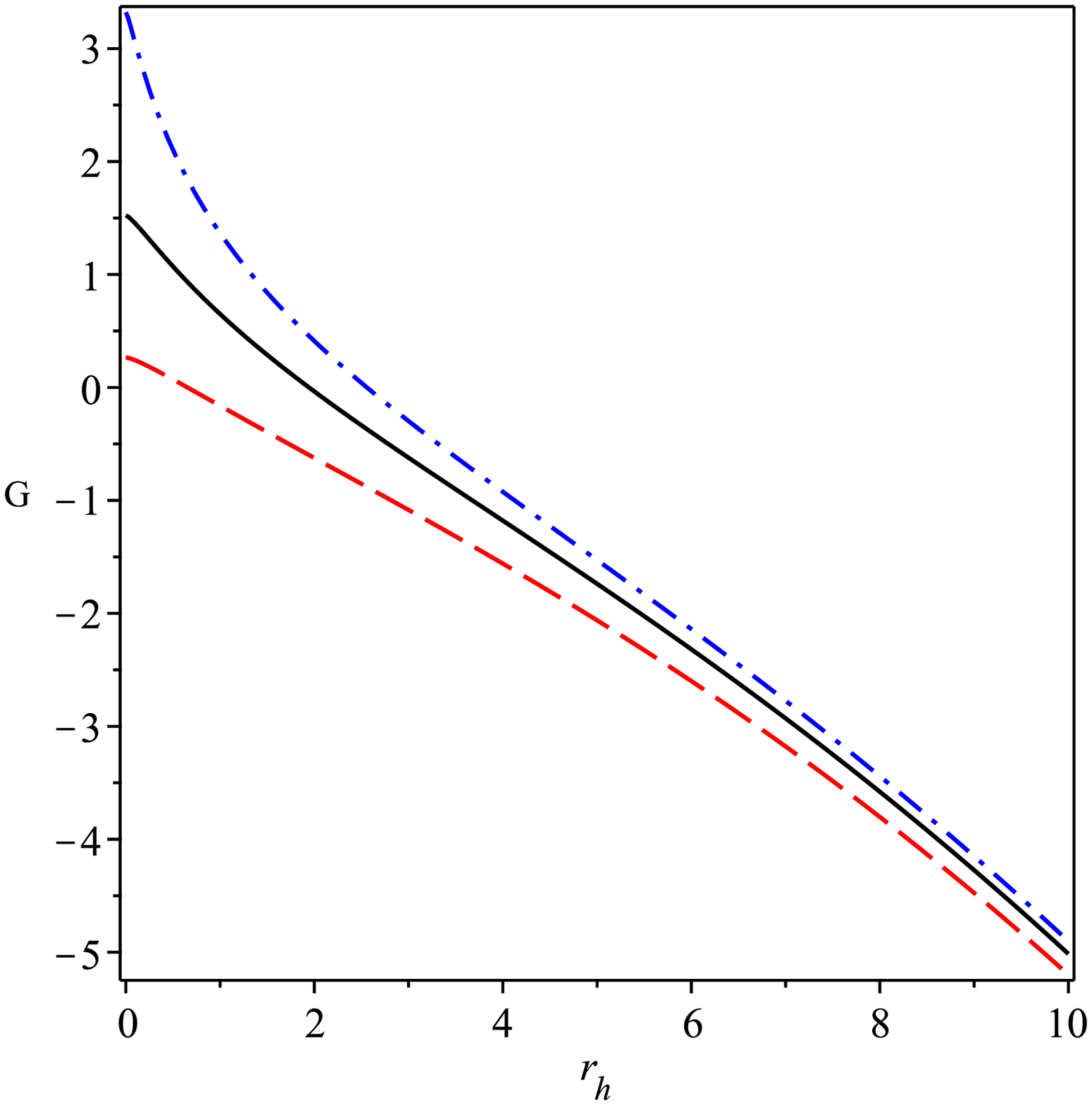}}\hfill
\subfloat[$Q=5$]{\includegraphics[width=.5\textwidth]{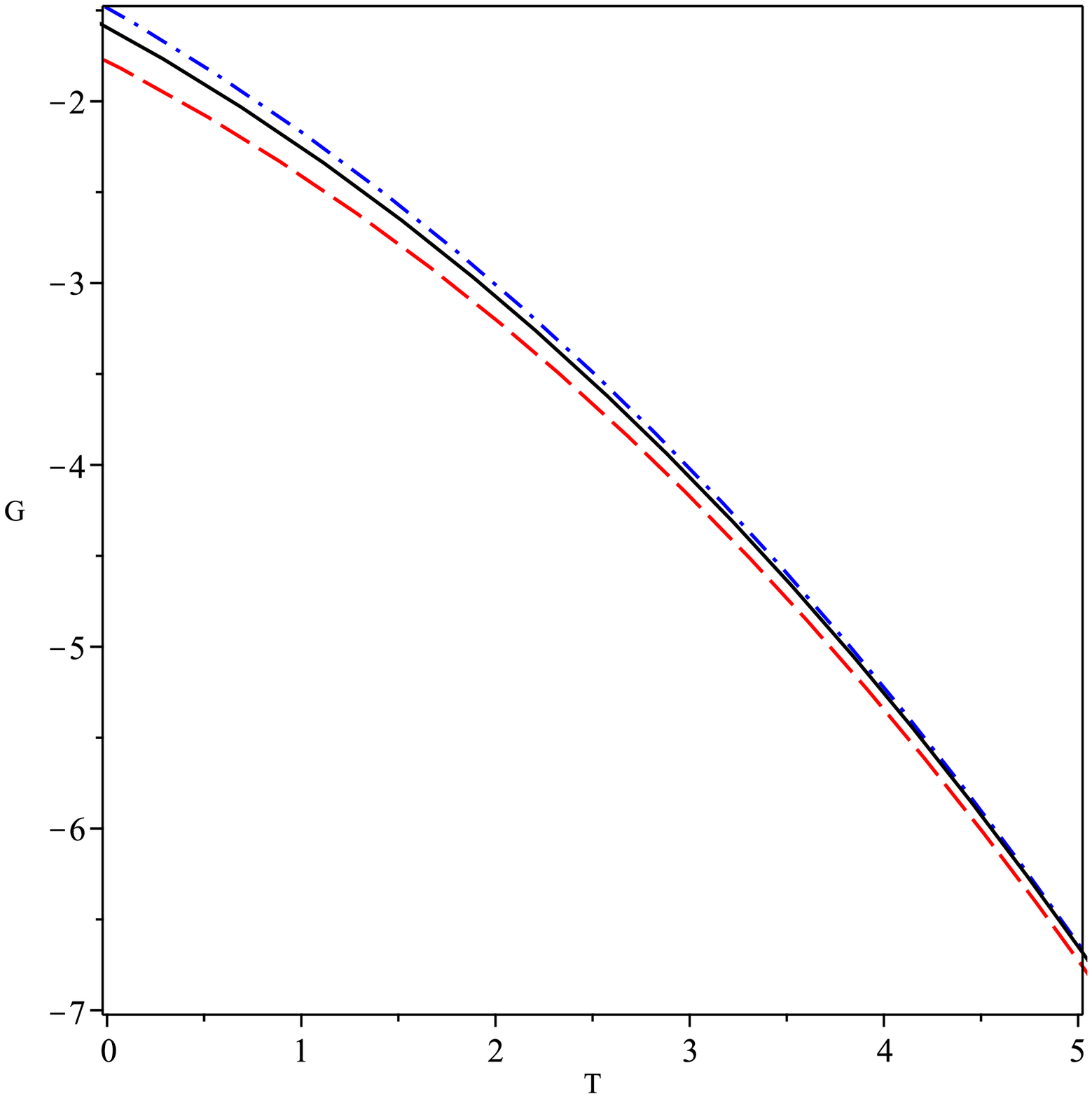}}\hfill
\caption{ Left panel: Gibbs free energy Vs horizon radius. $\beta=0.1$ denoted by black solid line, $\beta=0.5$ denoted by red dash line, $\beta=0.01$ denoted by blue dash-dot line with $l=2$. Right panel: Gibbs free energy Vs Temperature. $\beta=0.1$ denoted by black solid line, $\beta=0.5$ denoted by red dash line, $\beta=0.01$ denoted by blue dash-dot line with $l=2$.}\label{fig:6}
\end{figure}

The Gibbs free energy is depicted in Fig. \ref{fig:6}(a) and \ref{fig:6}(b) as a function of horizon radius and Hawking temperature. In Fig. \ref{fig:6}(a), Gibbs free energy started from a positive value, then it attained zero for some critical values of horizon radius. Finally, it goes to negative \& decreasing function of horizon radius. If we fixed the horizon radius then the Gibbs free energy of black holes represented by the red dash-dot line is more negative compared to the blue dash and black solid line, i.e. when NED parameter $\beta$ increases Gibbs free energy is more negative. The $G-T$ plot is depicted in Fig. \ref{fig:6}(b). The BTZ black hole does not show swallowtail-like behaviour, which indicates that no first-order phase transition occurs for such a black hole.

The Gibbs free energy in canonical ensemble is defined as $ G= \mathcal{M}- T_{H} S$, and using equations \eqref{eq:3.1}, \eqref{eq:3.2} and \eqref{eq:3.4} we obtain

\begin{equation}\label{eq:12}
G=\frac{\sqrt{r_{h}} \sqrt{Q a +r_{h}} Q}{8 a}-\frac{r_{h}^{\frac{3}{2}} \sqrt{Q a +r_{h}}}{4 a^{2}}-\frac{r_{h}^{2}}{8 l^{2}}+\frac{r_{h}^{2}}{4 a^{2}}-\frac{Q^{2}}{16} \ln  \Biggr[\frac{e (\frac{Q a}{2}+r_{h} +\sqrt{Q a r_{h} +r_{h}^{2}})}{2l}\Biggr].
\end{equation}

\begin{figure}[H]
    \centering
    \includegraphics[scale=0.5]{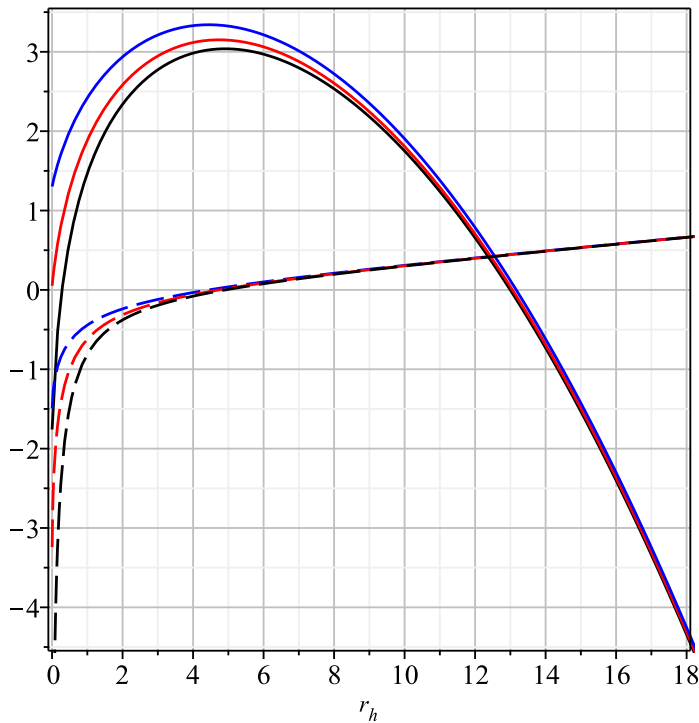}
    \caption{Gibbs free energy (in the grand canonical ensemble) \& Hawking temperature of the black hole Vs horizon radius. G (solid line) \& $T_{H}$ (dash line), Black colour represents $\beta=0.01$, Red colour represents $\beta=0.1$ \& Blue colour represent $\beta=0.5$ with $Q=5$ and $l=2$.}
    \label{fig:}
\end{figure}

The Gibbs free energy in the above equation is depicted in Fig. \ref{fig:6}. To study the global stability of the black hole thoroughly, we use the definition of Gibbs free energy in the grand canonical ensemble as Refs. \cite{zhang2019thermodynamical,dehghani2020nonlinearly}. Analyzing the signature of Gibbs free energy allows us to ascertain the global stability of black holes and identify the point of the Hawking-Page phase transition. According to the findings of Hawking and Page, a black hole can be considered globally stable if its Gibbs free energy exhibits a positive value. On the other hand, black holes that are unstable go through the Hawking-Page phase transition. In this context, any black hole characterized by negative free energy is assumed to be in the thermal or radiative phase \cite{hawking1983thermodynamics}. We plot the Gibbs free energy \& Hawking temperature in Fig. \ref{fig:} for different values of NED parameter $\beta$. Let $r_{h}^{min}$ \& $r_{0}$ be the zero's of Hawking temperature $T_{H}(r_{h}^{min})$=0 \& Gibbs free energy $G(r_{0})=0$ with $r_{h}^{min} < r_{0}$. Black holes that fall within the range of horizon radii $r_{h}^{min} < r_h < r_0$ exhibit global stability or black hole phase is preferable, while those with a horizon radius equal to $r_0$ undergo the Hawking-Page phase transition. Furthermore, black holes with horizon radii greater than $r_0$ tend to favour the radiative phase. It is worth noting that it follows from eq. \ref{eq:12} and Fig. \ref{fig:} that $\lim_{r_{h}\rightarrow 0}G\neq 0$ unlike the Einstein-AdS case Ref. \cite{hawking1983thermodynamics}. But at small radiuses, close to the Planck length
$l_P=\sqrt{G_N}$, quantum effects are important \cite{qm}. Therefore, without the quantum gravity theory, which is
not developed yet, the discussion of the case $r_h=0$ is questionable.

To analyze the Gibbs free energy, we differentiate Gibbs free energy as a function of charge and NED parameter $\beta$ of the black hole, while keeping the horizon radius fixed 
\begin{equation*}
\Bigl( \frac{\partial G}{\partial Q}\Bigl)_{r_{h},\beta}=0,
\end{equation*}
\begin{equation}
\Bigl( \frac{\partial G}{\partial \beta}\Bigl)_{r_{h},Q}=0.
\end{equation}
We find that Gibbs free energy is a monotonically decreasing function of charge, NED parameter $\beta$ \& nothing interesting behaviour (cusps, swallowtail) occurs.

To analyse the local stability \& phase transition of the black hole we study specific heat \& Hawking temperature. Specific heat of the black hole is defined as $  C_{\Phi}= T_{H} \bigl( dS/dT_{H} \bigl)$, and using equations \eqref{eq:3.1}, \eqref{eq:3.4} we have

\begin{equation}\label{eq:3.13}
C_{\Phi}= \frac{\pi   \Bigl(  \frac{2 r_{h}}{l^{2}}-\frac{4 r_{h}}{a^{2}}-\frac{2 Q}{a}+\frac{4  \sqrt{Q ar_{h} +r_{h}^2}}{a^{2}}   \Bigl) }{\Bigl( \frac{4}{l^{2}}-\frac{8}{a^{2}}+\frac{4 Q a +8 r_{h}}{a^{2}  \sqrt{Q ar_{h} +r_{h}^2}} \Bigl)}.
\end{equation}

\begin{figure}[H]
\centering
\subfloat[$Q=3$, $l=2$]{\includegraphics[width=.5\textwidth]{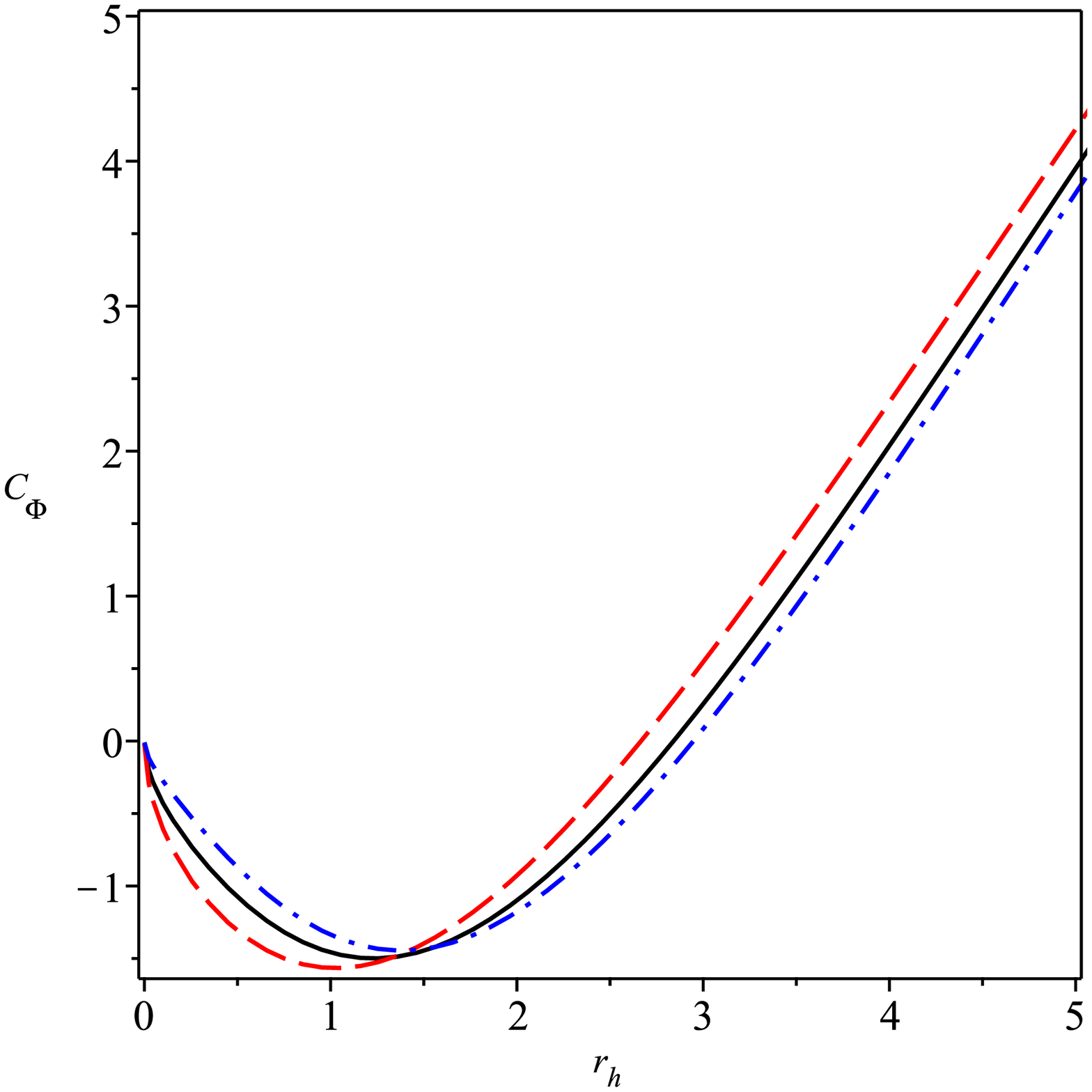}}\hfill
\subfloat[$\beta=1$, $l=2$]{\includegraphics[width=.5\textwidth]{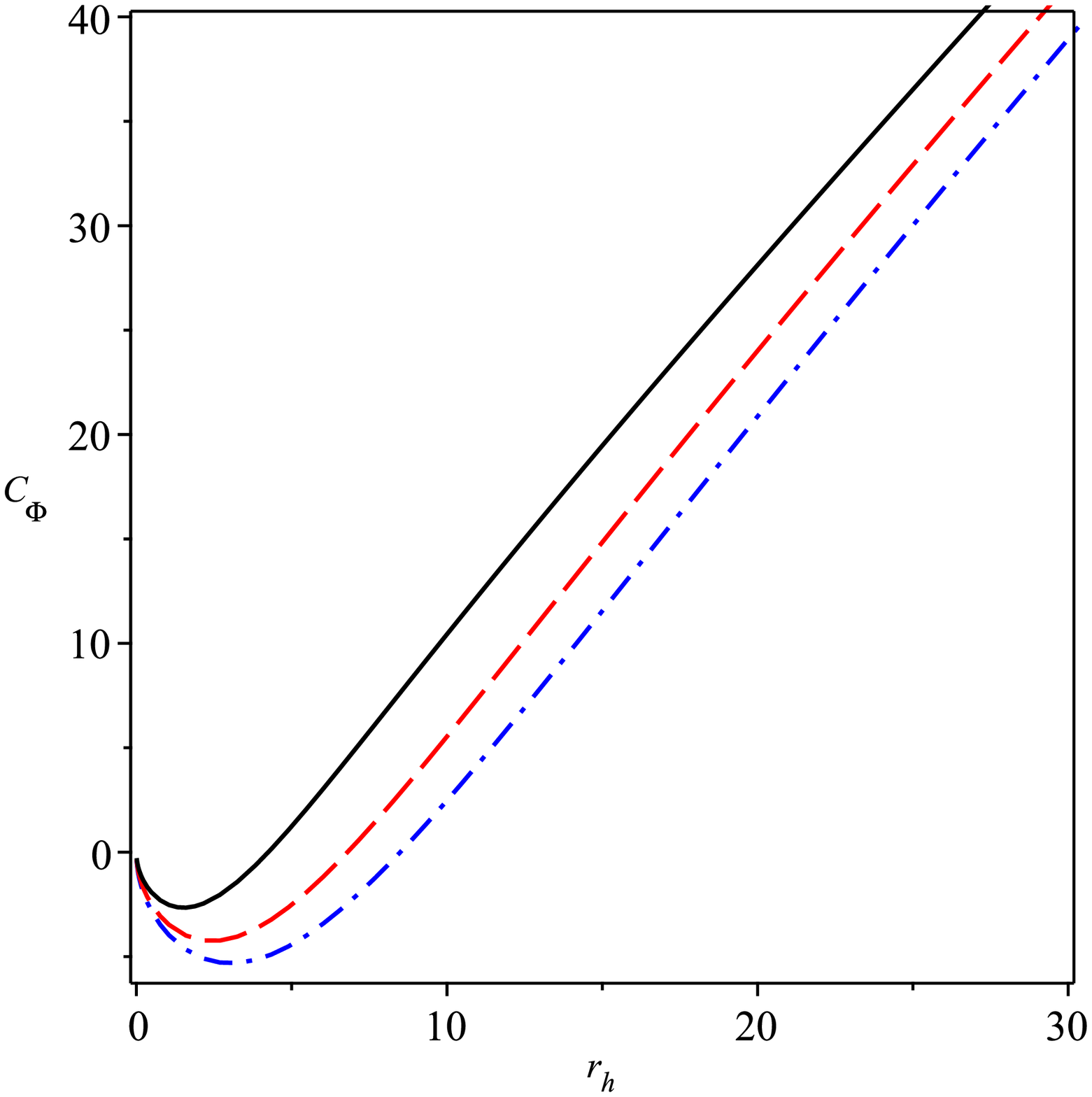}}\hfill
\caption{ Left panel: $\beta=0.1$ denoted by black solid line, $\beta=0.5$ denoted by red dash line and $\beta=0.01$ denoted by blue dash dot line.  Right panel: $Q=5$ denoted by black solid line, $Q=8$ denoted by red dash line and $Q=10$ denoted by blue dash-dot line.}\label{fig:7}
\end{figure}

In Fig. \ref{fig:7}(a) \& \ref{fig:7}(b), we plot the specific heat of black holes for different values of $\beta$ \& charges. For small-sized black holes specific heat is negative \& for massive-sized black holes specific heat is positive. The specific heat of the black hole is zero at $r_h=0$ and then it takes negative values. For a critical value of horizon radius specific heat is zero. The critical horizon radius can be obtained from the above equation by setting $C_{\Phi}=0$ and which is $r_{h}^{min}= Ql^2/(a+2l)$. Therefore, in the region $0 < r_{h} < r_{h}^{min}$ specific heat of the black hole is negative. If we examine the behaviour of the Hawking temperature in equation \eqref{eq:3.4}, it is found that in the region $0 \le r_{h} < r_{h}^{min}$ Hawking temperature takes negative values. At $r_{h}=r_{h}^{min}$ both Hawking temperature and specific heat of the black hole is zero. When $r_{h}>r_{h}^{min}$  both Hawking temperature and specific heat of the black hole are positive. In conclusion, black holes are locally thermodynamically stable when $r_{h} \ge r_{h}^{min}$ with positive temperature $(T_{H} \ge 0)$ \& specific heat $(C_{\Phi} \ge 0)$. The Gibbs free energy of the black hole in Fig. \ref{fig:6}  does not show swallowtail-like behaviour, therefore no first-order phase transition occurs for such black holes. The specific heat of the black hole is a continuous function of the horizon radius, which indicates that no second-order phase transition occurs for such a black hole.

\begin{figure}[H]
    \centering
    \includegraphics[scale=0.5]{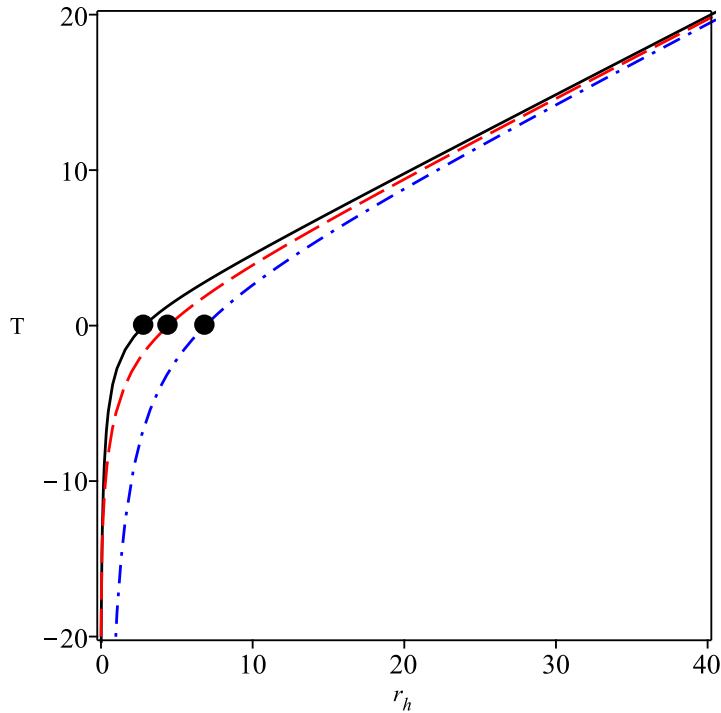}
    \caption{Hawking temperature of the black hole Vs horizon radius with $Q=3$ and $\beta=0.1$ denoted by black solid line, $Q=5$ and $\beta=0.5$ denoted by red dash line, $Q=7$ and $\beta=0.01$ denoted by blue dash-dot line.}
    \label{fig:8}
\end{figure}

In Fig. \ref{fig:8}, the Hawking temperature with respect to black hole horizon is depicted for different values of $\beta$ and charges of the black holes. From the figure it is clear that there is a minimum horizon radius, we mark this minimum radius by solid black circle. The temperature corresponding to minimum radius is zero when the radius of the black hole below the minimum radius temperature takes a negative value. As the negative temperature is not physical, therefore black holes can not exist below the minimum horizon radius. Therefore, the region $0 \le r_h<r_{h}^{min}$ is not physical region. The region where Hawking temperature is positive, i.e. $T_H \ge 0$ or $r_h \ge r_{h}^{min}$ is the physical region of parameter space. By setting $T_{H}=0$ into equation \eqref{eq:3.4} we obtain minimum horizon radius of the black holes 
\begin{equation}\label{eq:3.14}
    r_{h}^{min}= \frac{Q l^2}{a+2l}.
\end{equation}

To analyze the Hawking temperature very thoroughly, we differentiate Hawking temperature as a function of charge and NED parameter $\beta$ of the black hole, while keeping the horizon radius fixed
\begin{equation*}
    \Bigl(\frac{\partial T_{H}}{\partial Q}\Bigl)_{r_{h},\beta}=0,
\end{equation*}
\begin{equation}
    \Bigl(\frac{\partial T_{H}}{\partial \beta}\Bigl)_{r_{h},Q}=0.
\end{equation}
We find that Hawking temperature is a monotonically decreasing function of charge \& at $Q=0$ it attains a maximum. As a function of $\beta$ Hawking temperature is an increasing function of beta \& at $\beta=0$ it attains a minimum.

\section{Van der Waals black holes}\label{sec:4}

The Van der Walls equation of state  is

\begin{equation}\label{eq:4.1}
     \Bigl( P+\frac{a}{v^2} \Bigl) (v - b) = T,
\end{equation}

where $v$ is the specific volume of the fluid. The equation of state for a black hole obtained from equation \eqref{eq:3.4} is given by

\begin{equation}\label{eq:4.2}
P =\frac{T}{4 r_{h}}+\frac{Q}{8 \pi  a r_{h}}+\frac{1}{4 \pi  a^{2}}-\frac{\sqrt{Q a +r_{h}}}{4 \pi  a^{2} \sqrt{r_{h}}}.
\end{equation}

To obtain critical points we must set
\begin{equation}\label{eq:4.3}
 \biggl( \frac{\partial{P}}{\partial{r_{h}}}\biggl)_{T_{c}} =\biggl( \frac{\partial^{2}{P}}{\partial{r_{h}}^{2}}\biggl)_{T_{c} }=0, 
\end{equation}

and above equations have no solutions for $T_{c}>0$.

\section{Conclusions}\label{sec:6}
We have obtained a new electrically charged black hole solution in $(2+1)$ dimensional gravity coupled to NED in $AdS$ background. The metric and mass functions of black holes are studied and depicted in Fig. \ref{fig:1}. Keeping the charges of the black hole fixed, if we increase $\beta$ then the position of the horizon increases (Fig. \ref{fig:1}a) and similarly, if we increase electric charges then the position of the horizon also increases (Fig. \ref{fig:1}b). Next, we derive the first law of black hole thermodynamics in extended phase space. The cosmological constant plays the role of the thermodynamic pressure. Hawking temperature and mass of the black holes are computed. Using Hawking temperature and the mass of the black holes, various thermodynamics quantities (e.g. electrostatic potential, the Helmholtz and Gibbs free energies) are investigated. The global stability of black holes is studied through the analysis of Gibbs free energy \& Hawking temperature. It is found that black holes within the range of horizon radii $r_{h}^{min} < r_h < r_0$ are globally stable. To study the local stability of the black holes, specific heat was computed for different values of coupling $\beta$ and charges. The plot shows that smaller-sized black holes are unstable, as the size of the black holes increases specific heat takes a positive value, i.e. massive-sized black holes are more stable. From the $G-T$  plot, one can conclude that no first-order phase transition occurs for electrically charged black holes in $(2 + 1)D$ gravity coupled to NED. The plot of Hawking temperature shows that there is a minimum horizon radius below which a black hole can not exist. In section \ref{sec:4}, we study  Van der Waals-like phase transitions of black holes and it is found that no solutions are possible for $T_{c},v_{c} \ge 0$.

\printbibliography
\end{document}